# A Modified UDP for Federated Learning Packet Transmissions

Bright K. Mahembe, Clement Nyirenda

*Department of Computer Science*
*Faculty of Natural Science*
*University of the Western Cape*
*Cape Town 7532, South Africa*
kudzibright@gmail.com
cnyirenda@uwc.ac.za

*Abstract*—**This paper introduces a Modified User Datagram Protocol (UDP) for Federated Learning to ensure efficiency and reliability in the model parameter transport process, maximizing the potential of the Global model in each Federated Learning round. In developing and testing this protocol, the NS3 simulator is utilized to simulate the packet transport over the network and Google TensorFlow is used to create a custom Federated learning environment. In this preliminary implementation, the simulation contains three nodes where two nodes are client nodes, and one is a server node. The results obtained in this paper provide confidence in the capabilities of the protocol in the future of Federated Learning therefore, in future the Modified UDP will be tested on a larger Federated learning system with a TensorFlow model containing more parameters and a comparison between the traditional UDP protocol and the Modified UDP protocol will be simulated. Optimization of the Modified UDP will also be explored to improve efficiency while ensuring reliability.**

## I. Introduction

With the ever so growing amounts of dispersed big data being produced as a result of Artificial Intelligence (AI) driven technological advancements, motivated towards simplifying the day-to-day life of any individual. Efficiency, reliability, and security in the context of data have grown to be of great importance in the future success of these AI inventions. Researchers involved around the introduction and maintenance of these AI driven systems have introduced multiple frameworks and concepts to accommodate these complex systems. Federated Learning (FL), being one of those vital concepts that are believed to provide some level of comfort in ensuring data transfer, is efficient, reliable, and secure [1]. Federated learning was first introduced in 2016 by Google in an edge-server architecture designed to update language models on mobile phones while preserving data-privacy [1],[2]. Since then, Google TensorFlow and Pytorch have adapted their open-source platforms for developers to explore FL in-depth [3], [4].

The concept of FL includes a central server containing a global Machine Learning (ML) model communicating with multiple clients to orchestrate data exchange transactions in a round basis where in each round the global model goes through a data aggregation process that aggregates the given client model with the existing global model [5], [6]. With this exciting ML model enriching concept being introduced, the data exchange transaction protocols used to route machine learning model parameters between the client and the server in a FL architecture exists within the transport layer of the OSI model. Currently, the most commonly used transport layer protocols are Transmission Control Protocol (TCP) and the User Datagram Protocol (UDP). These protocols were not primarily designed for FL; therefore, they have limitations that would reduce the effectiveness of the FL approach.

TCP utilizes a connection-oriented approach which was designed to ensure high reliability, sacrificing the ability to quickly exchange data making it less efficient in the context of packet delivery per unit, which will minimize the learning potential of the model at a given time period [7]. On the other hand, UDP is a much simpler protocol which prioritizes the fast exchange of data; however, UDP provides no reliability resulting in a high possibility of packet loss which directly impacts the quality of the ML model after the training process [7]. Both efficiency and reliability are key areas in ensuring a successful FL orchestration. With the TCP and UDP protocols lacking in one of the two, there is a need to ensure that data arrives at the server and client in an efficient and reliable manner. Therefore, this paper proposes a modified UDP that utilizes the fast packet exchanges of the traditional UDP whilst ensuring the reliability of the transferred packets. The modified UDP protocol is developed in a Ubuntu environment using an NS3 simulator to simulate the network and TensorFlow to represent the FL aspect. The performance of the proposed transport protocol is evaluated in order to assess its ability to ensure no packets containing FL parameters are lost in each FL round. Different simulation scenarios are explored.

This paper continues with an in-depth description of the development of the proposed protocol split amongst the following sections. Section II describes FL fundamentals; section III provides an in-depth analysis of the functioning and logic around the proposed modified UDP. Section IV includes the experimental setting and obtained results, and lastly section V contains the conclusion and future work around the proposed concept.

## II. Federated Learning Fundamentals

In this section the fundamentals of FL are explored and the implementation of the FL architecture using NS3 is discussed. First, a comparison is made between the traditional centralized approach and the decentralized FL approach to provide context on the improvements made in ensuring data privacy is adhered to. As briefly described in section I, a FL architecture contains at least one client node and one server node. We will discuss each node's responsibilities in detail and proceed to the algorithm that performs the aggregation of parameters within the architecture.

## A. Server Node

The responsibilities of the server node within the FL architecture have significantly evolved as compared to the centralized approach. In FL, the server is responsible for the enhancement of the global model which represents the combined client ML parameters and is updated regularly to continuously improve various processes [4], [5]. Federated Averaging (FA) was introduced to grant the server the ability to perform ML parameter aggregation, this allows the server to collect ML model parameters from its pool of clients and perform some ML model aggregation computation that improves the performance of the global model [6], [7]. Once training is complete, the server sends the global model to its pool of clients as a form of an update as illustrated in Fig. 1 [8], [9].

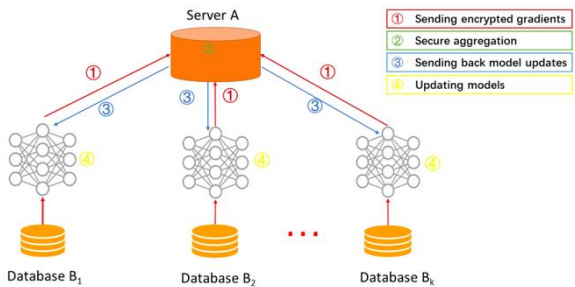

Fig. 1. Federated Learning Architecture. Adapted from [1]

## B. Client Node

In a centralized architecture the client nodes are solely responsible for the collection and transportation of raw data to a remote server for ML training [3] which raised some data privacy concerns. To eliminate the risk of potential data leakage, the client node evolved to perform its own analysis locally and ensure sensitive data does not get transported over the network [10]. In a FL architecture, the client nodes remain the primary source for the server; however, in this architecture the client nodes send ML parameters to the server instead of raw data [11]. Client nodes are responsible for the local training of data using a ML algorithm and routing the trained ML model parameters over a network to the server node for aggregation, this approach eliminates data privacy issues over the network [6], [8].

## III. IMPLEMENTATION OF FL IN NS3

In this research an NS3 simulator was adapted to simulate the data transmission functionalities needed to represent an end-to-end FL transaction. As part of the development of the Modified UDP for FL protocol, a star network topology, where N client nodes are served by one central server is developed. In the current implementation N is set to two in order to create a simple FL architecture. These nodes and their implementation, using Google TensorFlow and NS3 Simulator, are described as follows.

## A. Client Node Perspective

The client node begins by training a local model using an MNIST dataset of handwritten digit images through Google TensorFlow's Keras package and saves it locally for later use. When ready to send, the client node inputs the saved model, retrieves the weights and temporarily stores them in memory. Before sending each weight, the client converts the weight to a hexadecimal representation and proceeds to adding it into an NS3 packet that is transported through the network by the NS3 simulator to the server node that awaits the data. Algorithm illustrates the high-level view of the client's perspective using Google TensorFlow.

Algorithm I: Client Node Data Transportation Algorithm

```
1  Algorithm I: Client Node Data Transportation Algorithm.
2
3  INPUT Tensorflow_Model
4
5  weights <- Tensorflow_Model.get_weights()
6
7  for weight in weights Do
8      weightHexadecimal <- ConvertToHex(weight)
9      packet <- new packet()
10     packet.addPacketData(weightHexadecimal)
11     ns3.simulator.SendToServer(packet)
12
13 ListenForServerresponse()
14
15
```

## B. Server Node Perspective

When the server receives the client packets it begins its process by retrieving the data from the packet. The server proceeds to convert the digested data from a hexadecimal representation to its original form. Once all the packets are received, the server proceeds to aggregate the global model with the received parameters and send a packet containing the final response either successful or failed to the client node. See algorithm II for a high-level view of the implementation of the server node.

Algorithm II: Server Node Data Aggregation Algorithm

```
1  Algorithm II: Server Node Data Aggregation Algorithm.
2
3  INPUT ClientDataPacket
4
5  if(allPacketsReceived()):
6      globalModel <- Aggregation (GlobalModel,ReceivedParameters)
7      packet <- new Packet()
8      packet.addPacketData(Response)
9      ns3.sendToClient(packet)
10
11 ListenForClientData()
12
13
```

## C. Machine Learning Parameter Aggregation

The model aggregation procedure is what completes the data sharing transaction in a FL architecture, in this research we opted for a simple averaging aggregation technique. This takes two input parameters including the global model parameters and the client parameters to be aggregated into a new combined set of parameters, using

$$\sum_{i=0}^{N_{\text{Parameters}}} \frac{\text{Client}(i) + \text{Server}(i)}{2} \quad , \quad (1)$$

where *NParameters* denotes the total number of weights contained in the model starting at the first parameter which is 0 in this case. Corresponding client and server parameters are summed to produce a result that is divided by 2 to obtain the average weight that gets set within the new set of weights as illustrated in algorithm III.

Algorithm III: Parameter Aggregation

```
1   Algorithm III: Parameter Aggregation
2
3   INPUT Server Parameters, Client Parameters
4   OUTPUT AggregatedParameters
5
6   AggregatedParameters <- An array of the new set of Parameters
7
8   For x <- 0 , Total Number of Server Paremeters Do
9       For y <- X, length of the Weight found at Server Parameter index X Do
10          AggregatedParameters(X,Y) = (Server Parameters(X,Y) + Client Parameters(X,Y))/2.0
11
12  return AggregatedParameters
13
```

*D. Scalability*

The proposed NS3 based FL architecture can easily scale to N nodes in an FL orchestration where N is bounded by the computational capabilities of the hosting environment. Scaling up is as simple as indicating the number of nodes required for the simulation and the NS3 based FL framework will handle the networking, routing, and FL orchestration for you.

IV. A MODIFIED UDP FOR FEDERATED LEARNING

In this section we will discuss the steps taken to introduce reliability to the existing UDP for use in FL scenarios. This section begins by describing the current UDP process flow and proceeds to outlining the enhancements made towards ensuring reliability.

*A. TCP vs UDP*

The UDP was originally designed to transport data over a network for scenarios that are heavily dependent on fast transaction over reliability. The UDP protocol does not on its own ensure any reliability in recovering lost packets. This approach significantly reduces the transmission overhead of the UDP as compared to the reliable TCP [12][13]. Fig. 2. Illustrates the main differences between the UDP and TCP communication protocols.

The steps and processes taken by the UDP and TCP communication protocols in scheduling a successful data transmission process in their own capabilities are illustrated in Fig. 2. Before sharing data with the receiver, the TCP performs a 3-way handshake ensuring that the receiver is available before the data sharing transaction begins. This process increases the communication overhead of the TCP protocol making it a difficult choice for scenarios that prioritize speed. The UDP is a great protocol for those speed dependent scenarios, however, in the exception of scenarios such as an FL environment where efficiency and reliability are key, both protocols are not capable of delivering both.

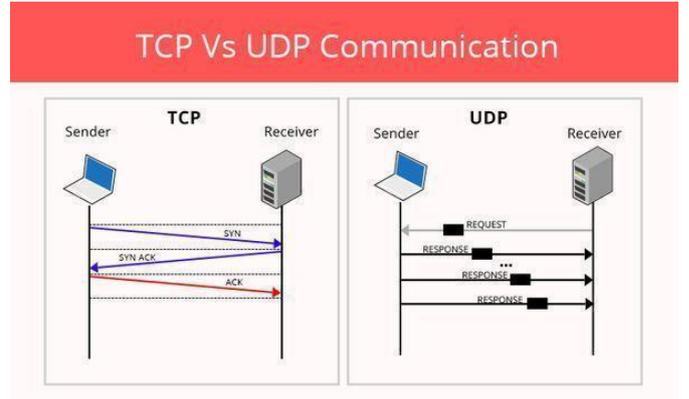

Fig. 2. TCP vs UDP comparison adapted from [14]

*B. A Modified UDP*

The modified UDP protocol aims to provide efficiency and reliability in one protocol using the UDP communication protocol as its base. The Modified UDP protocol performs the following steps in ensuring reliable data transmissions on the Sender as well as on the receiver.

A.  *Sender*
1.  Sends the packets as required in quick succession.
2.  Keep all the packets that have been sent for possible resending in case of packet losses.
3.  Starts a timer for determining when to resend the packets
    - If the sender receives an acknowledgement with sequence number (0, 0, A), then it means that all packets have been received and the transaction completes.
    - If the sender receives an acknowledgement with a sequence number lower than (X , Np; A) where is Np is the total number of packets, A is the address of the sender and X is the packet sequence number being a real number between $\{ 0 < X \leq N \}$
    - If no acknowledgement is received during the set period as tracked by the timer, it means there are missing packets the sender resends the last packets to inform the receiver of the missing sequences with Y amount of maximum retries.

B.  *Receiver*
1.  Receive all packets and properly store them.
2.  Once the last packet is received the receiver performs the following steps depending on the outcome:
    - If all packets have been received, then send an acknowledgement with sequence number (0, 0, A).
        - Construct the original file from the packets and proceed with the federated learning issues
        - Clear the storage locations
    - If some packets are missing, send acknowledgements with sequence numbers of only those missing packets. Start the timer for determining when to resend the acknowledgement.

The last packet sequence is used by the server to initiate the lost packet recovery procedure, in the event of the last packet (Np, Np, A) being lost, the sender resends that packet to signal the receiver to respond with the sequence numbers of the missing packets. Fig. 3 provides a visual illustration of a simple transaction between the sender and the receiver as described above.

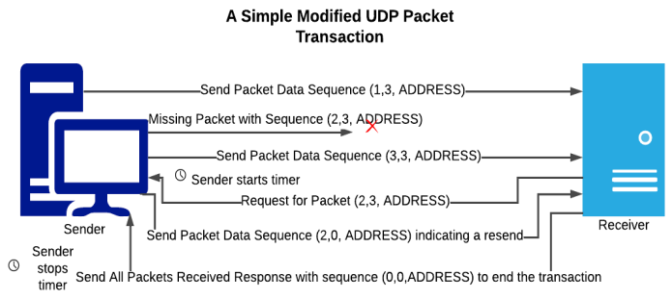

Fig. 3. A simple Modified UDP transaction.

## V. PERFORMANCE EVALUATION

In this section, we will run a simulation of the modified UDP protocol within an FL architecture over one round. Fig. 4 provides a visual illustration of a single FL round using the Modified UDP approach. The client retrieves data from the database and trains it in a TensorFlow model. Once the model is trained the client sends all the model weights to the server in the form of packets and begins the response timer.

If the server does not receive all the packets, it will respond with the packets it needs referencing them by their sequence number. If the server does not respond within the client's time period, the client resends the last packet to the server with a maximum of 3 retries to get it to request any missing packets. If the server receives all packets, it responds with a sequence of (0, 0, A) informing the client to stop the timer and end the transaction.

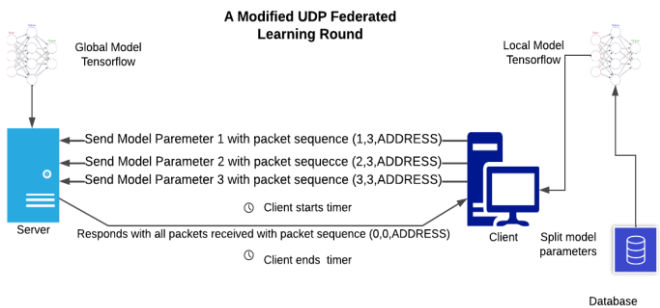

Fig. 4. A single round of the Modified UDP protocol in an FL Scenario.

### A. Simulation environment

The simulation is performed on an Intel(R) Core(TM) i3 8th generation laptop with 8gb of ram running Linux Ubuntu OS. NS 3.30 was used to create the simulation network in a 3 node connection with a data Rate of 5mbps and a delay of 2000ms. In this simulation we train a small TensorFlow model with at most 4 packets on the client and transport the trained parameters within the packets to the server over the network.

### B. Test cases

Two test cases are implemented. In both cases, the four FL packets are transmitted. In both cases, the server retrieves all of the missing packets to complete the training for the specified client in the FL round.

The client deliberately skips packet (2, 4, 10.1.2.4). Figure 5 shows the process in the Linux test environment. In this test, the following IPV4 addresses are assigned to the nodes; 10.1.2.4 is assigned to Client node one, and 10.1.2.5 is assigned to the Server node.

```
Agent 1 preparing to send
1875/1875 [==============================] - 5s 2ms/step - loss: 0.2979 - accuracy: 0.9134
File saved
Timer Started

Now at Packet 1 of 4
1
Now at Packet 3 of 4
3
Now at Packet 4 of 4

Server attempting to retrieve lost packet : 2

Packet: 2 is missing!
Agent preparing to send missing packet: 2

Now at Packet 2 of 2

Transaction Complete Global Model aggregated +17511535993.0ns
Model reference <keras.engine.sequential.Sequential object at 0x7fb701a8cb50>
Server Replying

All packets Received on the Server for Agent:<network.InetSocketAddress object at 0x7fb6fe9997f0>
Timer Stopped
```

Fig. 5. Modified UDP FL round missing packet (2, 4, 10.1.2.4)

In Test case 1 visually illustrated in Fig. 5, the client trains the local model and saves the weights locally. The client proceeds to transport the trained weights in packets to the server which logs the packet sequence it receives to the terminal. When the last packet is received by the server, the recovery process begins. The server contacts the client via the network to resend the missing packet (2, 4, 10.1.2.4). The client will respond with the packet and awaits for the server's response packet sequence (0, 0, 10.1.2.5) to conclude the transaction.

### C. Test case 2

The client deliberately skips packets (2, 4, 10.1.2.4); (3, 4, 10.1.2.4); and (4, 4, 10.1.2.4). Unlike test case 1, in test case 2 the last packet is missing so the server does not initiate the recovery process, the client timer expires in the process and the client resends the last packet in its history packet sequence (4, 4, 10.1.2.4). When the server receives a packet sequence (4, 4, 10.1.2.4) it attempts to recover the missing packets (2, 4, 10.1.2.4) and (3, 4, 10.1.2.4) in a similar fashion as in test case 1. When the clients are received by the server the FL round concludes with the server sending a response with the packet sequence (0, 0, 10.1.2.5). Fig. 6 illustrates this scenario.

### D. Test case 3

Client two simulates a successful transaction where all packets are delivered to the server and the timer stops when the server responds to avoid transaction delays for other clients as visually illustrated in Fig. 7.

```
Agent 1 preparing to send
1875/1875 [==============================] - 4s 2ms/step - loss: 0.2978 - accuracy: 0.9141
File saved
Timer Started

Now at Packet 1 of 4
1
Timer has ended without a response, re-sending last packet
Agent preparing to send missing packet: 4

Now at Packet 4 of 4

Server attempting to retrieve lost packet : 2

Packet: 2 is missing!
Agent preparing to send missing packet: 2

Now at Packet 2 of 2

Server attempting to retrieve lost packet : 3

Packet: 3 is missing!
Agent preparing to send missing packet: 3

Now at Packet 3 of 3

Transaction Complete Global Model aggregated +17511535993.0ns
Model reference <keras.engine.sequential.Sequential object at 0x7f5cba71fb50>

Server Replying

All packets Received on the Server for Agent:<network.InetSocketAddress object at 0x7f5cb762c810>
Timer Stopped
```

Fig. 6. A Modified UDP FL round missing packets (2 ,4, 10.1.2.4); (3, 4, 10.1.2.4); and (4, 4, 10.1.2.4).

```
Agent 2 preparing to send
1875/1875 [==============================] - 4s 2ms/step - loss: 0.2990 - accuracy: 0.9121
File saved
Received +20513183992.0ns

Now at Packet 1 of 4
Sending +21013183992.0ns
1
Received +25516479990.0ns

Now at Packet 2 of 4
Sending +26016479990.0ns
2
Received +30519775988.0ns

Now at Packet 3 of 4
Sending +31019775988.0ns
3
Received +35523071986.0ns

Now at Packet 4 of 4

Transaction Complete Global Model aggregated +35523071986.0ns
Model reference <keras.engine.sequential.Sequential object at 0x7faf379b4d00>

Server Replying
Sending +36023071986.0ns

All packets Received on the Server for Agent 2
Timer Stopped
```

Fig. 7. Client two transaction

## VI. CONCLUSION AND FUTURE IMPROVEMENTS

This paper introduced a Modified UDP protocol for FL that provides efficiency while ensuring sufficient reliability motivated towards maximizing the data enriching potential of the global model in each FL round. The Modified UDP results provide confidence in the capabilities of the protocol in the future of FL and potential cases where efficiency and reliability are essential. During the course of testing we were able to retrieve all missing data and ensure the global model was trained to its highest potential.

Further research and tests will be conducted on improving various aspects of the proposed Modified UDP protocol. In future research we will evaluate the performance differences of the traditional UDP protocol against the proposed Modified UDP in an FL orchestration. We will add more clients to the FL round, and send larger model parameters to measure the impact the protocol has on the overall performance of the TensorFlow ML model.

**Bright K. Mahembe** is a Software Engineer at CyberIAM. He obtained his Honours Degree in Computer Science at the University of the Western Cape and is currently pursuing his MSc degree in Computer Science at the University of the Western Cape. His research interests are in Artificial Intelligence, Networking and Cyber Security and their application in Communication and process automation.

**Clement N. Nyirenda** received his PhD in Computational Intelligence from Tokyo Institute of Technology in 2011. His research interests are in Computational Intelligence paradigms such as Fuzzy Logic, Swarm Intelligence, and Artificial Neural Networks and their applications in Communications.